\documentclass[twocolumn,showpacs,preprintnumbers,amsmath,amssymb]{revtex4-1}
\usepackage{graphicx}
\usepackage{dcolumn}
\usepackage{bm}
\usepackage[latin1]{inputenc}
\usepackage{amsfonts}
\usepackage{amssymb}
\usepackage{amsmath}
\begin{document}



\title{Spin-Strain Phase Diagram of Defective Graphene}
\author{E.~J.~G.~Santos}
\affiliation{
Centro de F\'{\i}sica de Materiales CFM-MPC,
Centro Mixto CSIC-UPV/EHU, Paseo Manuel de Lardizabal 5,
20080 San Sebasti\'an, Spain}
\affiliation{Donostia International Physics Center (DIPC),
Paseo Manuel de Lardizabal 4, 20018 San Sebasti\'an, Spain}

\author{S. Riikonen}
\affiliation{Laboratory of Physical Chemistry, Department of Chemistry, University of Helsinki
P.O. Box 55, FI-00014, Finland}

\author{D. S\'anchez-Portal}
\affiliation{
Centro de F\'{\i}sica de Materiales CFM-MPC,
Centro Mixto CSIC-UPV/EHU,  Paseo Manuel de Lardizabal 5,
20080 San Sebasti\'an, Spain}
\affiliation{Donostia International Physics Center (DIPC),
Paseo Manuel de Lardizabal 4, 20018 San Sebasti\'an, Spain}

\author{A. Ayuela}
\affiliation{
Centro de F\'{\i}sica de Materiales  CFM-MPC,
Centro Mixto CSIC-UPV/EHU,  Paseo Manuel de Lardizabal 5,
20080 San Sebasti\'an, Spain}
\affiliation{Donostia International Physics Center (DIPC),
Paseo Manuel de Lardizabal 4, 20018 San Sebasti\'an, Spain}


\begin{abstract}
Using  calculations on  defective graphene  from first  principles, we
herein consider the dependence of the properties of the monovacancy of
graphene  under isotropic  strain,  with a  particular  focus on  spin
moments.   At  zero  strain,  the  vacancy  shows  a  spin  moment  of
1.5~$\mu_B$ that increases to  $\sim$2~$\mu_B$ when the graphene is in
tension. The changes are more  dramatic under compression, in that the
vacancy  becomes non-magnetic  when graphene  is compressed  more than
2\%.  This  transition is  linked to changes  in the  atomic structure
that occurs around vacancies, and  is associated with the formation of
ripples.  For  compressions slightly greater than 3\%, this rippling
leads to  the formation of  a heavily reconstructed  vacancy structure
that  consists of two  deformed hexagons  and pentagons.   Our results
suggest that any defect-induced  magnetism that occurs in graphene can
be  controlled  by  applying   a  strain,  or  some  other  mechanical
deformations.
\end{abstract}


\maketitle

Since its discovery\cite{geim07}, graphene  has been shown to possess a
number    of   remarkable    electronic    and   elastic    properties
\cite{novoselov05}.  Its  electronic properties are  mainly associated
with the  $p_z$ orbitals of  carbon, while its elastic  properties are
associated with the $sp$ orbitals.  Specifically, a Young's modulus of
1  TPa\cite{wei08}  has  recently   been  measured  for  graphene,  in
agreement with the average Young\'\  s modulus for carbon nanotubes of
Y =1.25  TPa \cite{krishnan98,overney}.  Graphene  can sustain elastic
deformations as large  as 20\%, and it is typically  under a strain of
several  percent  when deposited  on  surfaces \cite{huanga09}.   This
strain  could  be  used  to  engineer  the  electronic  properties  of
graphene.  Unfortunately, a very high  tensile force must be applied to
defect-free graphene  to modify the electronic  structure, because the
spectrum   of   planar   graphene   remains  metallic   up   to   10\%
stretching\cite{gui08,pereira09,  colombo}.   However,  there are  not
many  experimental conditions  under which  layers of  graphene
remain strictly  planar.  Due  to the soft  transversal phonons  of 2D
graphene, at  certain temperatures the  layer shrinks and  ripples are
formed.  These ripples are present in free-standing layers of pristine
graphene\cite{fasolino07,meyer07}   and   for   layers  deposited   on
substrates. \cite{stolyarova07,ishigami07}.   The rippling of graphene
induces midgap states \cite{guinea08} in the electronic structure that
disappear  when relaxations  in the  layers are  carefully  taken into
account \cite{wehling08}.  The relaxations that accompany the rippling
of the  layer, and  their influence on  the electronic  structure, are
particularly  important in  layers that  are  under a  high degree  of
compression.  Strong rippling  of this  kind  can also  be induced  by
adsorbates  \cite{Car08}  and defects  \cite{Bangert09}  such as  that
recently discovered  for OH impurities \cite{thompson09,meyer07,kern}.
Understanding the interaction  between defect-induced deformations and
ripples is  one of the main challenges  now faced in the  study of the
electronic structure of graphene.

In  pristine graphene,  the  $sp$ contributions  are  situated at  low
energies, and it is the  presence of defects that promotes them closer
to the  Fermi level.  By theoretical calculations,  it has  been shown
that   substitutional  doping\cite{santos08,santos10a,santos10b,lee97}
and the presence of defects\cite{lehtinen,lehtinen03} in graphene, and
in  graphitic  materials in  general,  produce  magnetism  that is  of
interest in the  potential use of these materials  in spintronics.  In
experiments,                      the                      irradiation
\cite{esquinazi03,ohldag07,krasheninnikov07,gomez05}      and      ion
bombardment  \cite{ugeda} of  carbon-based materials  create vacancies
which  indeed  are  linked   with  magnetic  signals.   The  vacancies
\cite{pereira06,lehtinen,yazyev98}  and  edges \cite{fujita96,enoki07}
present in graphene layers have been the focus of detailed theoretical
studies.  Although there have been  both studies of the changes in the
electronic  structure induced  by rippling  and  studies of
defects in  graphene, to our knowledge  there have been  no studies of
the  magnetism  of rippled  graphene.   Such  a study  is of  particular
interest  when  considered  alongside  the  effect of  strain  on  the
structural and electronic properties of graphene.

We  herein  combine  the  two  perspectives and  present  a  study  of
defective graphene  under strain.   We focus on  a particular  type of
defect, namely  related to carbon  monovacancies, and find  that these
show  a rich  diversity of  structural and  spin phase-behavior  as a
function of strain.  For zero strain, each vacancy has a solution with
spin  moment of 1.5~$\mu_B$.   However, non-magnetic  solutions become
stable under  a moderate  compression of less  than 2\%  strain.  This
transition in magnetic moment may be traced to the modification in the
local atomic structure of the  defect as ripples develop in the layer.
In  calculations  for  compressions  greater  than  3\%,  the  vacancy
structure  departs from its  usual geometry  under rippling,  and this
reconstruction ends with the formation of a different defect structure
that has a  central atom with strong $sp^3$  hybridization.  All these
structural changes  are linked with  the appearance of ripples  as the
layer   is   compressed,  similar   to   those   observed  in   recent
experiments~\cite{ripples07,Vazquez08,Morgenstern09}.     We    herein
explore the relationship between  the appearance of rippling patterns,
the presence of vacancies, and  changes in the electronic and magnetic
properties of graphene.



We   perform   density   functional
calculations~\cite{kohn1965}  from first  principles using  the SIESTA
code~\cite{soler02}.      We    use    the     generalized    gradient
approximation~\cite{gga},                               norm-conserving
pseudopotentials~\cite{troullier91}  and  a  basis  set  of  numerical
atomic orbitals  ~\cite{soler02}.  The integration  over the Brillouin
zone        uses         a        well-converged        Monkhorst-Pack
$k$-sampling~\cite{monkhosrt76}  that  is  equivalent to  $70\times70$
points for  the graphene unit cell.  The  structural optimizations use
the conjugate gradient algorithm,  which causes the forces to converge
until they are lower than 0.05 eV/\AA.


\begin{figure}[h]
\includegraphics[width=3.100in]{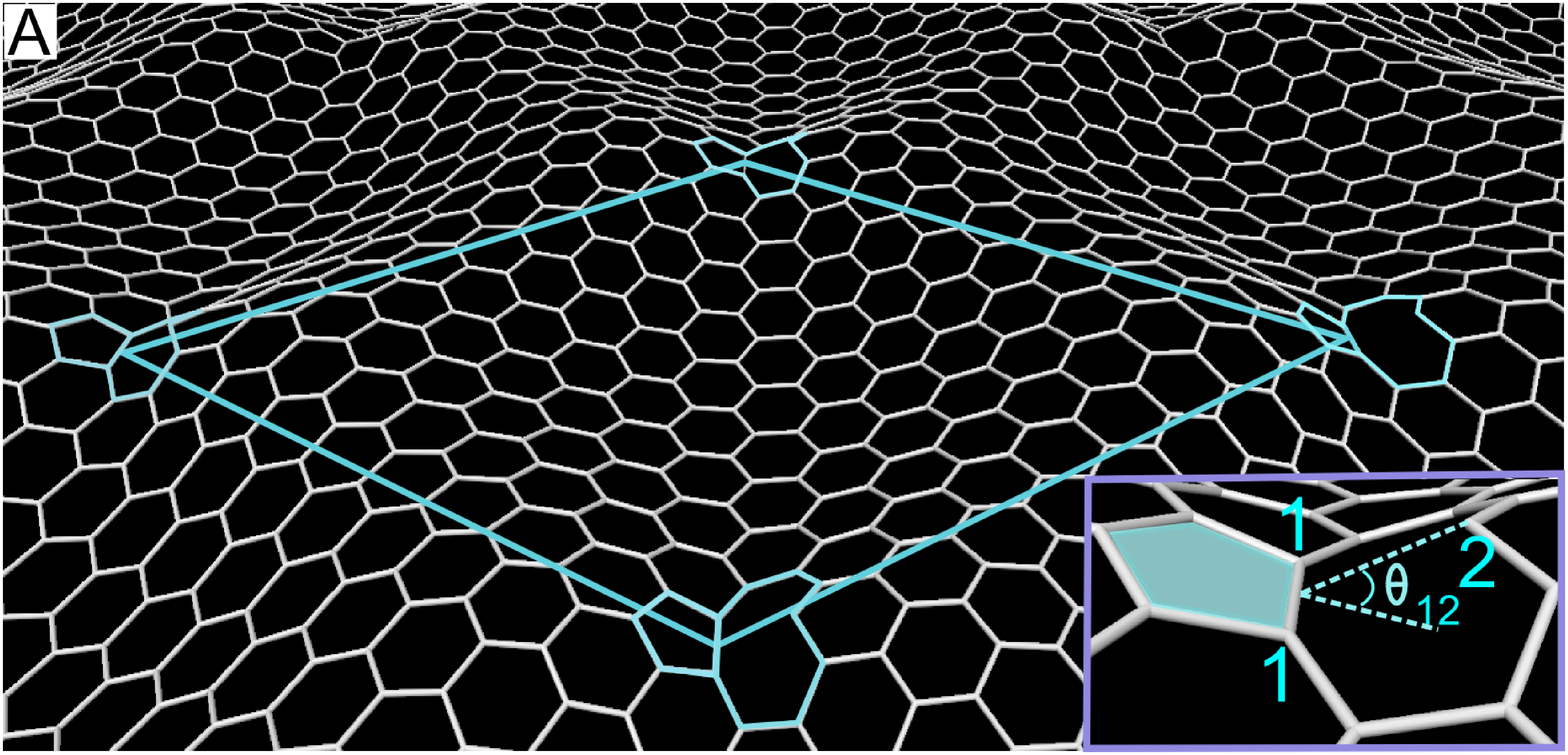}
\includegraphics[width=3.100in]{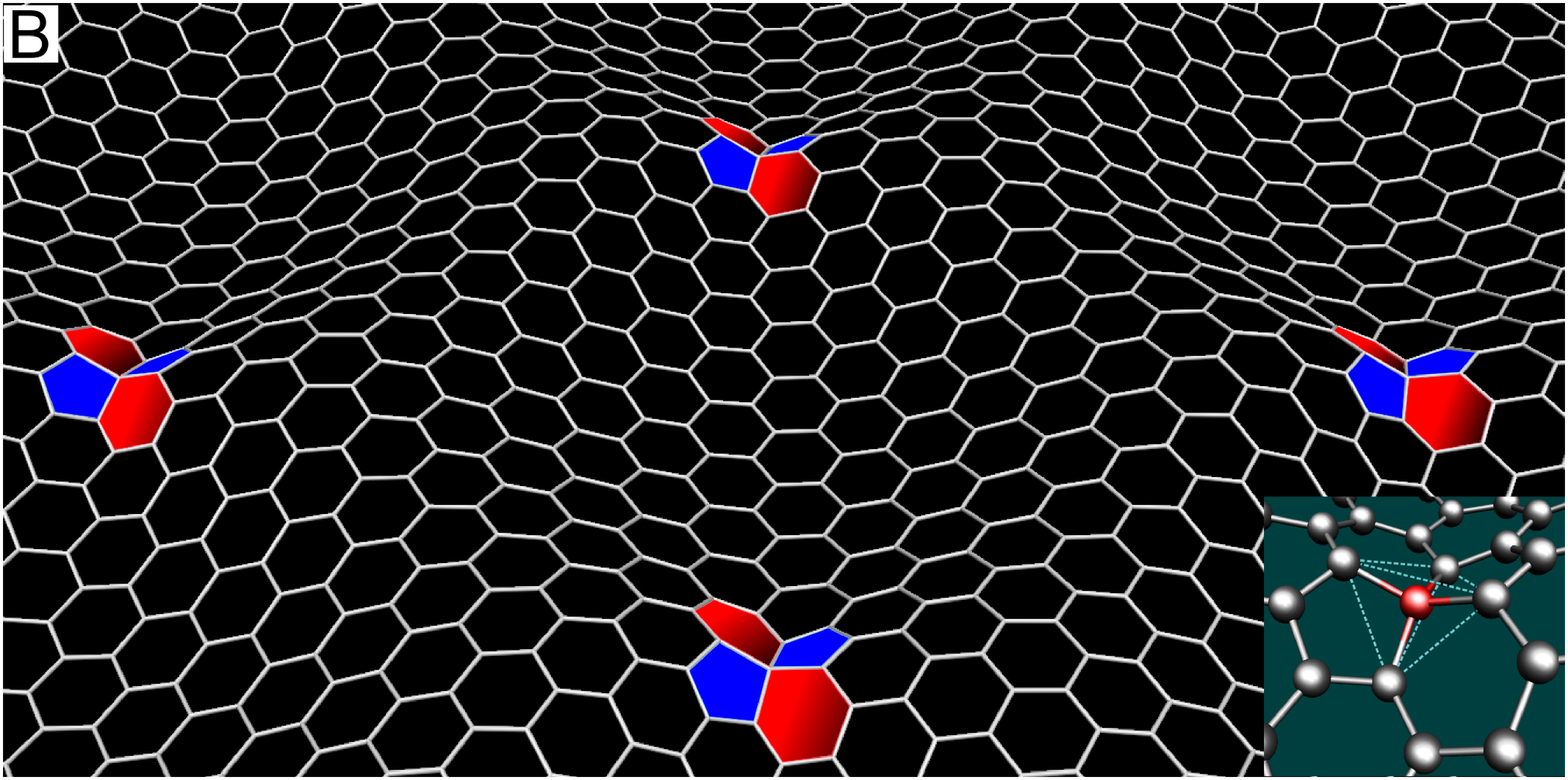}
\caption{  (A) Graphene  with vacancies  under an
isotropic compression of 1.2\%.  As an example we use the 10$\times$10
unit cell (highlighted). The inset shows the geometry of the vacancies
and  the  atomic  labels.   The  local bending  at  the  vacancies  is
described  by the  angle $\theta_{12}$  between the  pentagon  and the
plane  defined  by  three  C  atoms  around  the  vacancy,  i.e.   two
equivalent atoms  labeled 1  and a third  atom labeled 2.   Note the
rippling of graphene sheets with  vacancies at the saddle points.  (B)
A different vacancy structure for a compression slightly under 3\%. It
has two  distorted hexagons  and pentagons, while  the central  C atom
shows strong $sp^3$ hybridization. }
\label{fig1}
\end{figure}

{\it Structure  versus strain.} While under tension  the layer remains
perfectly  flat,  and   small  compressions  produce  the  spontaneous
rippling  of graphene  with  a characteristic  deformation around  the
vacancy.   \ref{fig1} illustrates  the  rippled geometry  after
relaxing  one of  our models  for a  free-standing  defective graphene
layer  under an isotropic  compression of 1.2\%.   At zero  strain, the
monovacancy  tends  to  undergo  a Jahn-Teller--like  distortion  that
lowers its energy by $\sim$200~meV: atoms  of type 1 (see the inset of
\ref{fig1}) reconstruct  to form a pentagon  with the neighboring
atoms, while  atom 2 is left with the dangling  bond responsible  for the
spin polarization.  Although we herein consider a non-planar structure
under strain, our findings at  low compressions are similar to results
previously   obtained  for   flat   graphene  ~\cite{amara07,ewels03}.
However, atom 2 in our  rippled structure is progressively lifted from
the graphene surface by as much as $\sim$1.0~\AA\ for strains slightly
below  $3\%$,   just  before  the  occurrence  of   a  strong  vacancy
reconstruction.   The data of  \ref{fig1}B reveal  a particularly
striking result. In contrast to 2D vacancy, for which previous authors
invariably   described  a   'pentagon-goggles'   structure,  our   own
calculations  show that  the vacancy  is strongly  reconstructed under
rippling. For compressive strains greater than 3\%, a different defect
structure is produced altogether, consisting of two heavily distorted
hexagons  and pentagons  and  resembling the  transition  states of  a
planar vacancy movement. We shall now focus on the description of this
region of transition for compression of up to 2.8\%.

\begin{figure}[h]
\includegraphics[width=2.25000in]{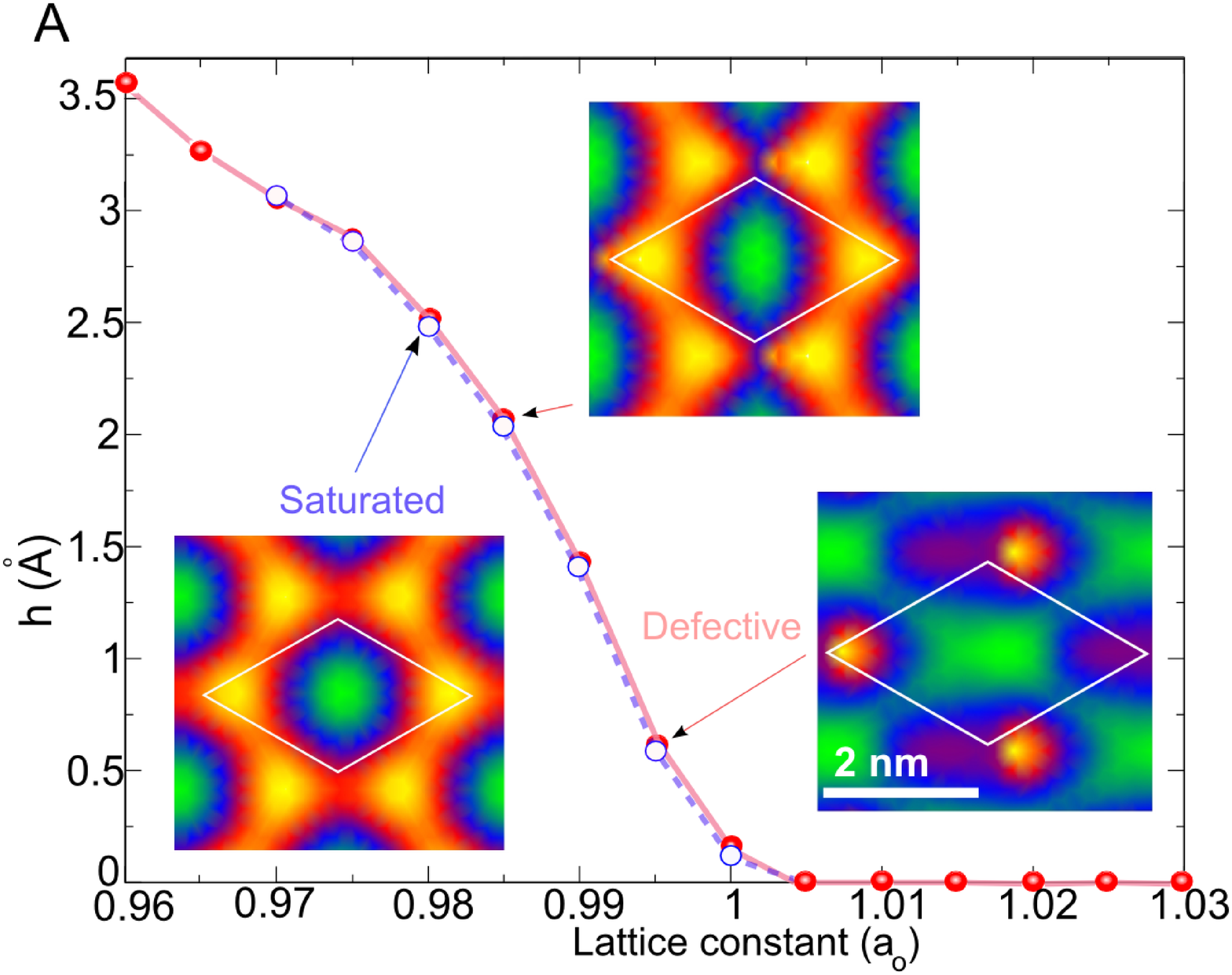}
\includegraphics[width=2.25000in]{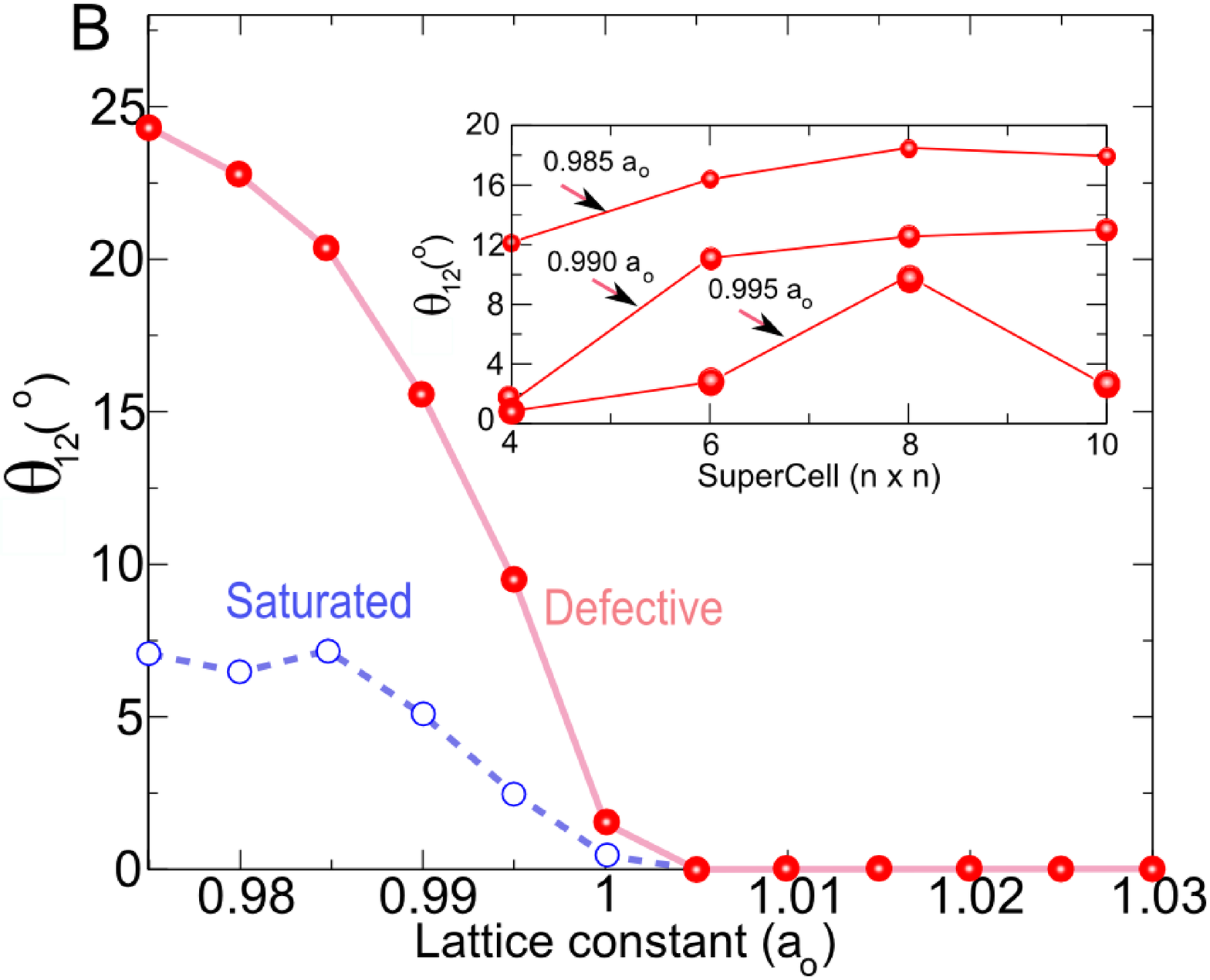}
\caption{(A) Maximum amplitude $h$(\AA)
of  rippling for  10$\times$10  graphene supercells  versus strain  in
units of the equilibrium  lattice constant $a_{o}$. The filled circles
denote defective graphene, while the empty circles denote the pristine
layer that has been recovered by  adding a carbon atom to the vacancy.
The   insets  show   the  corrugation   patterns:   the  bright/yellow
(dark/green) areas indicate  the higher (lower) regions.  (B)
Bending angle  $\theta_{12}$ (see definition in  \ref{fig1}) as a
function of strain. The inset shows the behavior of the bending angles
with  size of  supercell for  different values  of strain.   The local
induced deformation  clearly depends on the presence  of vacancies and
on the size of the supercell.  }
\label{fig2}
\end{figure}

\ref{fig2}A shows the maximum  amplitude of the rippling $h$
as a  function of  strain for defective  graphene (shown by  the solid
symbols).   We removed  the  defects from  each  rippled structure  by
adding carbon atoms  to the vacancy sites and  relaxing the structure.
We obtained rippled pristine graphene with maximum amplitudes as shown
in \ref{fig2} (open symbols).  The amplitudes and the topographic
patterns  (insets  in   \ref{fig2}A)  show  little  difference
between pristine  or defective graphene,  particularly for compressive
strains  greater  than  $\sim$1\%.   It  seems  that  even  for  large
supercells, the main determinant to get ripples is the applied
strain.   However, the  structural patterns  that occurs  in defective
graphene  define  a  preferential  direction  and  break  the  up-down
symmetry  that   exits  perpendicular  to  the  layer.    It  is  this
preferential  direction  and  the  breaks  in the  symmetry  that  are
explained by  the reconstructed  vacancies, which break  the hexagonal
symmetry  of  graphene  with  their  goggles-pentagon structure. Vacancies thus play a  key role in determining the shape and
symmetry of the global deformation patterns of graphene.

We simultaneously  characterize the  local curvature at  the vacancies
and the height of atom 2 above the local tangent-plane in terms of the
bending  angle  $\theta_{12}$,  as  defined  in  \ref{fig1}.   In
\ref{fig2}B, we plot the  bending angle for both defective and
pristine  graphene.   Under moderate  compression  (1-2\%), the  local
bending angle for defective graphene is about three times greater than
that of  pristine graphene and  the graphene rippling is  clearly made
easier  by   the  presence  of  the  vacancies.    The  dependence  of
$\theta_{12}$ on the  size of the supercell, as shown  in the inset of
\ref{fig2}B,  further demonstrates  the  coupling between  the
local geometry of  the defect and the global  deformation, in that for
$4\times4$ supercell, a strain larger than 1.0\% is required to obtain
an increase in $\theta_{12}$ and to allow the corrugation of the layer
to begin.  However, for  larger supercells, much smaller strains cause
appreciable deformations around the vacancies.

\begin{figure}[h]
\includegraphics[width=2.100000in]{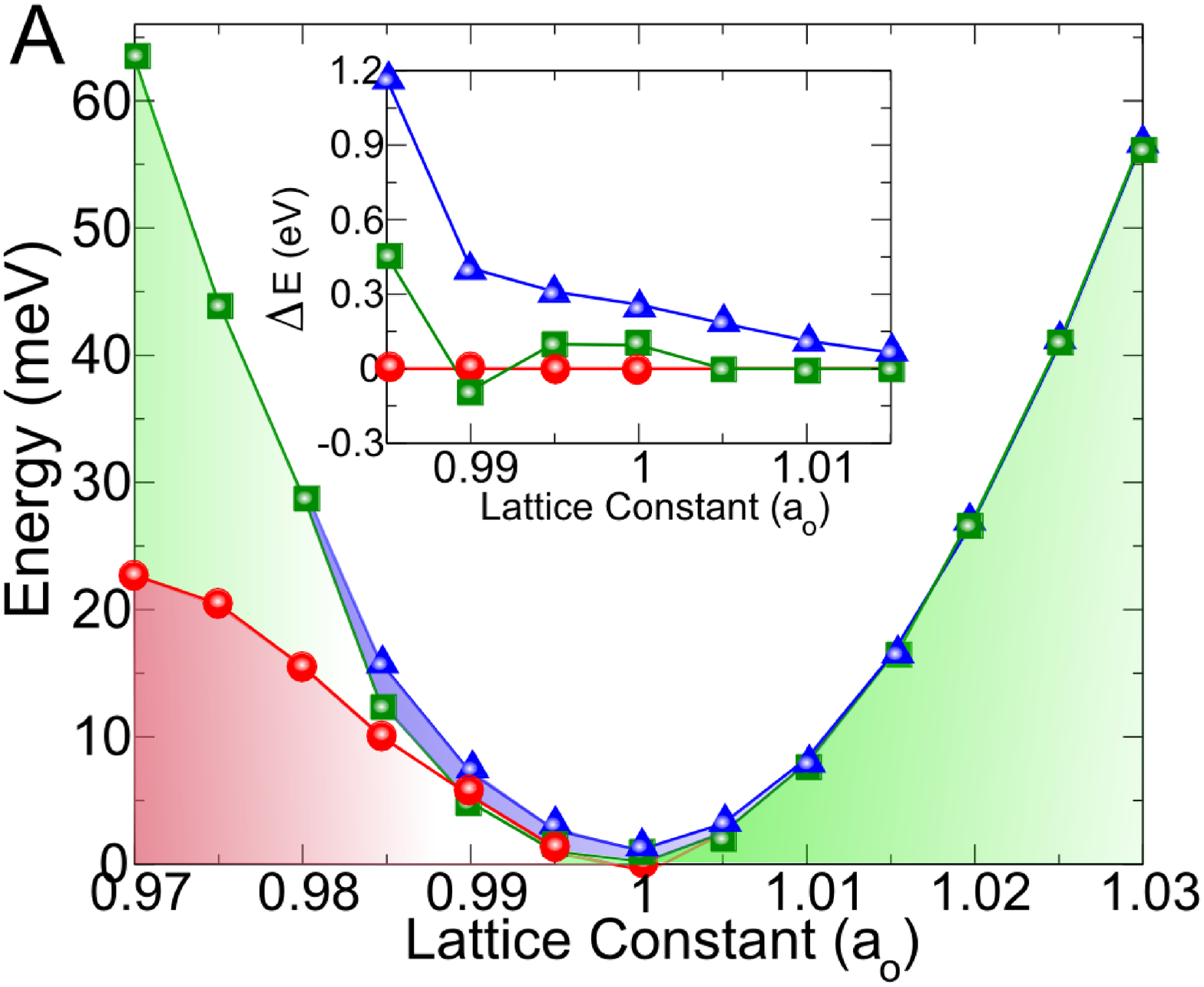}
\includegraphics[width=2.1000000in]{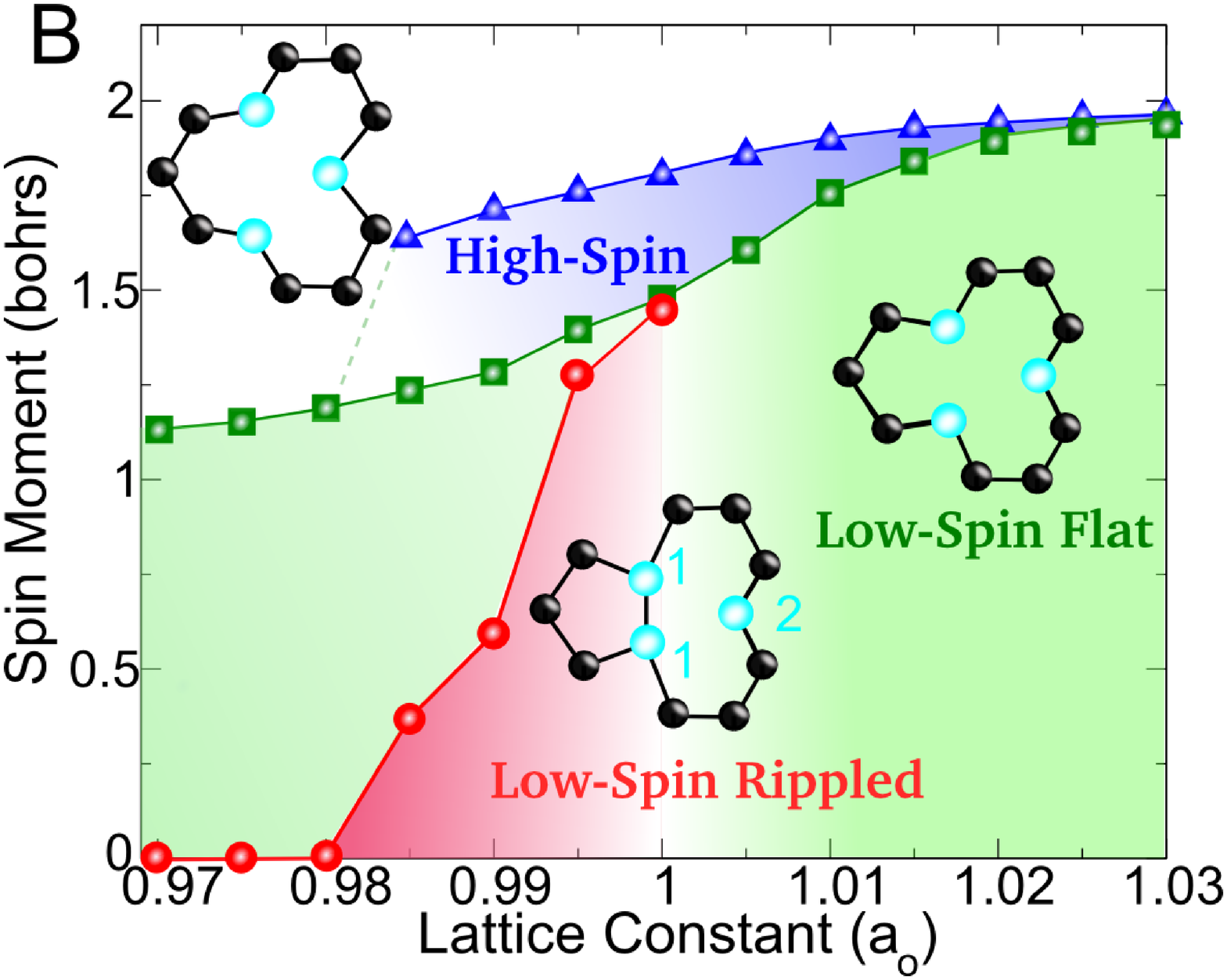}
\includegraphics[width=2.10000in]{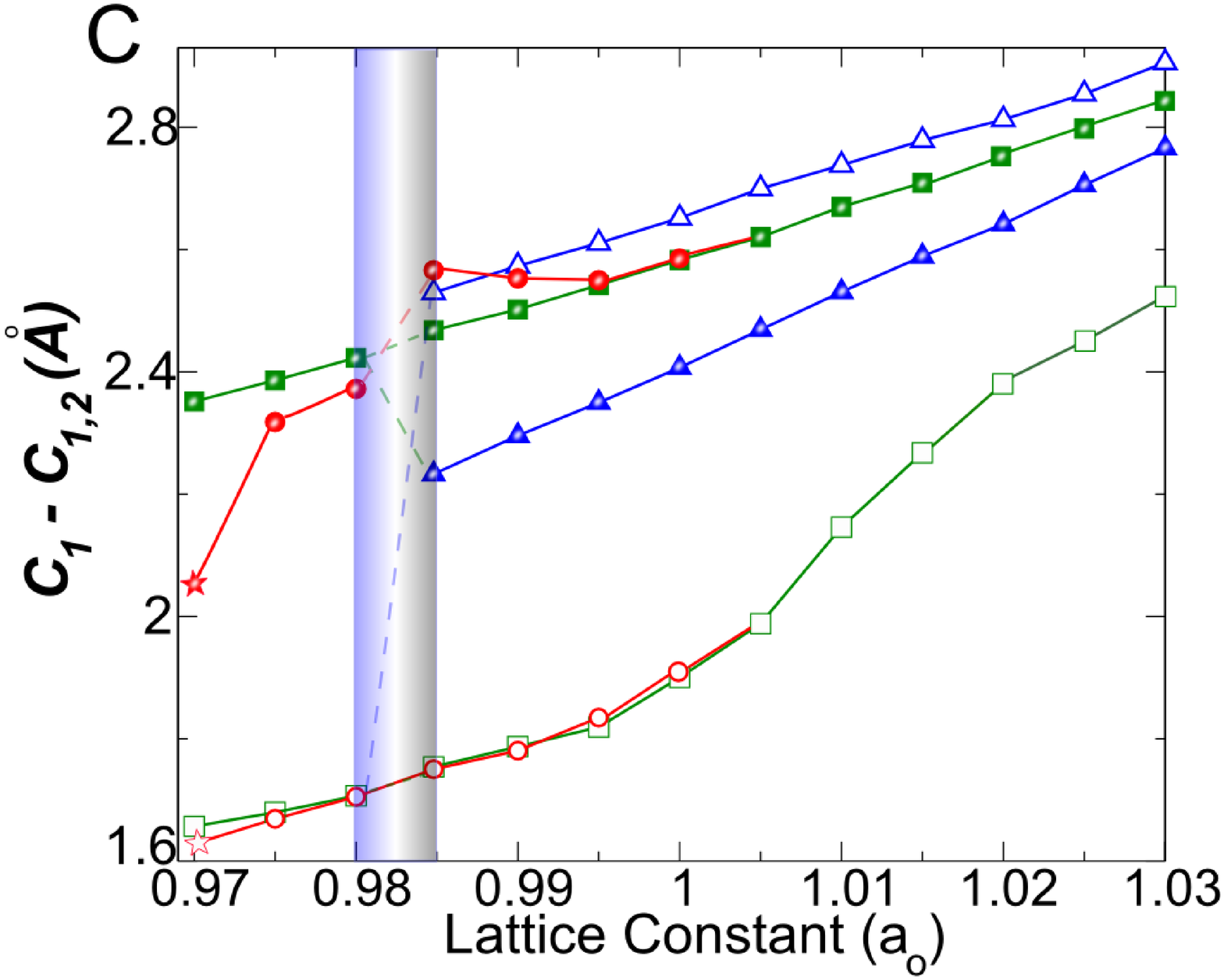}
\caption{Variation  in total energy per atom (A),
spin moment (B), and bond length (C) versus strain for three different
configurations  of vacancy  in the  10$\times$10  supercell: high-spin
flat  (triangles),  low-spin   flat  (squares)  and  low-spin  rippled
(circles).  The inset in panel  (a) details the change in total energy
with  respect to  the low-spin  solutions at  about zero  strain.  The
filled  and empty  symbols  in panel  (C)  represent the  1-2 and  1-1
distances, respectively  (see the inset of  \ref{fig1} for atomic
labels).  The marks  given by stars refer to  the average distances as
the  geometry departs  from  the  usual vacancy  to  the structure  in
\ref{fig1}B.  Note  that the magnetism disappears at  a strain of
$\sim$2\% when allowing if out-of-plane deformations are allowed.}
\label{fig3}
\end{figure}

{\it  Energetics}.  The  range  of  applied strain  used  here (a  few
percent)  is  comparable to  that  used  in  experiments in  which  an
appreciable corrugation of graphene  was reported for supported layers
\cite{strainAPL,Stroscio07,Vazquez08},    chemically    functionalized
graphene~\cite{Car08}  , or  defective  layers~\cite{Bangert09}. As  a
result of  the imposed  periodicity and the  finite-sizes used  in our
calculations,  long-wavelength  deformations  appear primarily  to  be
related  to the  strain,  i.e.  vacancies  do  not create  significant
corrugation  in  the  relaxed  geometry  at  the  equilibrium  lattice
constant.  However, we observe that for the range of strain considered
here,  the  system  is   compressed  and  thereby  assumes  a  rippled
configuration  at an  energy between  two  and three  times lower  for
defective graphene than for the pristine layer. For a 10x10 supercell,
the  energy required to  create ripples  for a  compression of  3\% is
reduced to almost  half its value in the  presence of vacancies, as seen
in \ref{fig3}A.  This difference in energy is consistent with the 
proposed role of vacancies as a source of ripples in recent 
experiments~\cite{Bangert09}.

{\it  Magnetism  versus  strain:  several  spin  solutions}.   In  the
foregoing part we  commented on the fact that  isotropic strain can be
used to tune  the vacancy structure and the curvature  of the layer at
the  site of  the defect.   We  now focus  on the  influence of  these
structural changes  on the magnetic  and electronic properties  of the
vacancies.  We  first consider  the case of  zero strain, i.e.  at the
equilibrium    lattice    constant.     The    usual    reconstruction
\cite{amara07,ewels03}  is  accompanied  by  a  decrease  in  the  1-1
distance in \ref{fig1} and an increase in the 1-2 distances.  The
two dangling bonds in the type 1 atoms are thus saturated while atom 2
remains uncoordinated.   It is  the polarization of  the corresponding
dangling bond that is the main  reason behind the appearance of a spin
moment  of $\sim$1.5$\mu_B$  associated with  the  carbon monovacancy.
However,   we   were   able   to  stabilize   another   reconstruction
characterized by  a different structural distortion, in  which the 1-2
distance decreases,  while the 1-1  distance increases [see  the local
structures  and distances  in \ref{fig3}B,C].   This structure
has  a larger  spin moment  of  $\sim$1.82$\mu_B$.  We  refer to  this
latter  structure as  the  high-spin (HS)  configuration,  and to  the
former  as the  low-spin (LS)  configuration.  At  zero strain  the LS
structure is more stable than the HS by around 250~meV.

The  behavior of  these  two structures  as  a function  of strain  is
summarized in  \ref{fig3}.  When applying  tension to the
layer, both  the HS  and LS structures  remain flat and  almost become
degenerate.  In  both cases the  spin moments show a  slight increase.
Conversely,  when the  layer is  compressed, the  HS  structure becomes
unstable.   Indeed,  it  is   only  possible   to  stabilize   the  HS
configuration under compression provided that the layer is constrained
to remain flat.  Even for flat graphene, the energy difference between
the HS and LS states increases significantly with compression, and for
strains  greater   than  1.5\%  the   HS  configuration  spontaneously
transforms into  LS-flat (i.e. low-spin  constrained to be  flat).  In
both  structures,   the  spin  moments   decrease  as  the   layer  is
compressed. The change is greater for the LS-flat configuration, which
varies from  1.98~$\mu_B$ for a $+$3\% deformation  to 1.15~$\mu_B$ at
$-$3\%.  A more  dramatic  reduction in  the  spin moment  is seen  if
ripples    are   allowed    to   form    in   the    graphene   layer.
\ref{fig3}B  shows that  the spin  moment of  the LS-rippled
(i.e.  low-spin free  to ripple)  vacancy decreases  sharply  when the
compressions  exceeds  0.5\%.   In  fact,  the  ground  state  of  the
monovacancy becomes non-magnetic for  compressive strains in the range
of 1.5-2\%.

The  changes in the  spin moments  of vacancies  may be  understood by
analyzing  their   electronic  structure.   When  carbon   atom  2  is
restricted  to  remain  in  the  plane,  the  spin  moment  is  mainly
associated with  the $sp^2$ dangling  bond, and the  contribution from
the $p_z$  states is smaller.   The strain-induced deformation  of the
vacancy and subsequent out-of-plane  displacement of atom 2 gives rise
to  the hybridization between  the out-of-plane  $p_z$ states  and the
in-plane $sp^2$ states.  For flat graphene the spin-polarized impurity
level  associated  with the  vacancy  therefore  has  a strong  $sp^2$
character  and  remains  essentially  decoupled from  the  delocalized
electronic levels of graphene  due to their different symmetries.  For
the rippled layer,  these two types of electronic  states are strongly
hybridized.  This  hybridization results in the  delocalization of the
defect  levels which  eliminate  the magnetism  and  explains why  the
LS-rippled configuration has a  lower spin moment.  Expressed in three
dimensions, the rippling is believed to transform the hybridization of
the C  vacancy atoms  in $sp^3$, thereby  removing the cause  of local
magnetism,  as  seen  at  the   higher  compression  of  the  case  of
\ref{fig1}B.

Rippling of  the graphene layer occurs  in a range of  cases. One such
case is where graphene is deposited on substrates, in which there is a
significant  mismatch  in between  lattice  parameters.  For  example,
ripples  have  been  previously  observed for  graphene  deposited  on
Ru(0001)~\cite{Vazquez08},  as well as  on other  metallic substrates.
Strain  can  also be  applied  and  controlled  by placing  exfoliated
graphene       on      flexible      substrates,       as      reported
recently~\cite{Mohiuddin09,Kim09}.  Under  such conditions, it  may be
possible to create  defects in regions of different  curvature using a
focused  electron  beam,  by  applying  a technique  similar  to  that
recently demonstrated by Rodriguez-Manzo and Banhart~\cite{Banhart09}.
The existence  and spatial distribution of the  spin moment associated
with the vacancies, and its variations with deformation, could then in
principle  be  monitored  using  a spin-polarized  scanning  tunneling
microscope.~\cite{Wiesendanger09,ugeda}

In summary, we have shown  that the magnetic and structural properties
of carbon vacancies depend strongly  on the local curvature induced by
defects.  Together  with the global rippling  pattern, this curvature
can  be controlled  by the  application of  a strain.   By compressing
graphene by up to 2\%, we can adjust the spin moment of the vacancy to
between  0  and  1.5~$\mu_B$.   At  our  high-compression  limit,  the
rippling allows the  defect C atoms to be arranged  in a geometry that
is different from the vacancy as generally observed.  Such strains may
be achieved experimentally \cite{Mohiuddin09,Kim09} and are comparable
with those  found when graphene is  deposited on a  range of different
substrates.  In fact,  our results suggest that the  magnetism that is
induced in graphene by the presence of defects can be controlled using
isotropic strain and other mechanical deformations.

{\it Acknowledgment.} We acknowledge the support of the Basque Departamento de Educaci\'on and the
UPV/EHU (Grant No. IT-366-07), the Spanish Ministerio de Innovaci\'on,
Ciencia y Tecnolog\'{\i}a 
(Grant  No. FIS2007-66711-C02-02),  and  the ETORTEK  research
program funded  by  the  Basque
Departamento de Industria and the Diputac\'ion Foral de Guipuzcoa.

\bibliography{draft12.bib}

\begin{thebibliography}{50}%
\makeatletter
\providecommand \@ifxundefined [1]{%
 \@ifx{#1\undefined}
}%
\providecommand \@ifnum [1]{%
 \ifnum #1\expandafter \@firstoftwo
 \else \expandafter \@secondoftwo
 \fi
}%
\providecommand \@ifx [1]{%
 \ifx #1\expandafter \@firstoftwo
 \else \expandafter \@secondoftwo
 \fi
}%
\providecommand \natexlab [1]{#1}%
\providecommand \enquote  [1]{``#1''}%
\providecommand \bibnamefont  [1]{#1}%
\providecommand \bibfnamefont [1]{#1}%
\providecommand \citenamefont [1]{#1}%
\providecommand \href@noop [0]{\@secondoftwo}%
\providecommand \href [0]{\begingroup \@sanitize@url \@href}%
\providecommand \@href[1]{\@@startlink{#1}\@@href}%
\providecommand \@@href[1]{\endgroup#1\@@endlink}%
\providecommand \@sanitize@url [0]{\catcode `\\12\catcode `\$12\catcode
  `\&12\catcode `\#12\catcode `\^12\catcode `\_12\catcode `\%12\relax}%
\providecommand \@@startlink[1]{}%
\providecommand \@@endlink[0]{}%
\providecommand \url  [0]{\begingroup\@sanitize@url \@url }%
\providecommand \@url [1]{\endgroup\@href {#1}{\urlprefix }}%
\providecommand \urlprefix  [0]{URL }%
\providecommand \Eprint [0]{\href }%
\providecommand \doibase [0]{http://dx.doi.org/}%
\providecommand \selectlanguage [0]{\@gobble}%
\providecommand \bibinfo  [0]{\@secondoftwo}%
\providecommand \bibfield  [0]{\@secondoftwo}%
\providecommand \translation [1]{[#1]}%
\providecommand \BibitemOpen [0]{}%
\providecommand \bibitemStop [0]{}%
\providecommand \bibitemNoStop [0]{.\EOS\space}%
\providecommand \EOS [0]{\spacefactor3000\relax}%
\providecommand \BibitemShut  [1]{\csname bibitem#1\endcsname}%
\let\auto@bib@innerbib\@empty
\bibitem [{\citenamefont {Geim}\ and\ \citenamefont
  {Novoselov}(2007)}]{geim07}%
  \BibitemOpen
  \bibfield  {author} {\bibinfo {author} {\bibfnamefont {A.~K.}\ \bibnamefont
  {Geim}}\ and\ \bibinfo {author} {\bibfnamefont {K.~S.}\ \bibnamefont
  {Novoselov}},\ }\href@noop {} {\bibfield  {journal} {\bibinfo  {journal}
  {Nat. Mater.}\ }\textbf {\bibinfo {volume} {6}},\ \bibinfo {pages} {183}
  (\bibinfo {year} {2007})}\BibitemShut {NoStop}%
\bibitem [{\citenamefont {Novoselov}\ and\ \citenamefont
  {other}(2005)}]{novoselov05}%
  \BibitemOpen
  \bibfield  {author} {\bibinfo {author} {\bibfnamefont {K.~S.}\ \bibnamefont
  {Novoselov}}\ and\ \bibinfo {author} {\bibnamefont {other}},\ }\href@noop {}
  {\bibfield  {journal} {\bibinfo  {journal} {Proc. Natl. Acad. Sci. U.S.A.}\
  }\textbf {\bibinfo {volume} {102}},\ \bibinfo {pages} {10451} (\bibinfo
  {year} {2005})}\BibitemShut {NoStop}%
\bibitem [{\citenamefont {Lee}\ \emph {et~al.}(2008)\citenamefont {Lee} \emph
  {et~al.}}]{wei08}%
  \BibitemOpen
  \bibfield  {author} {\bibinfo {author} {\bibfnamefont {C.}~\bibnamefont
  {Lee}} \emph {et~al.},\ }\href@noop {} {\bibfield  {journal} {\bibinfo
  {journal} {Science}\ }\textbf {\bibinfo {volume} {321}},\ \bibinfo {pages}
  {385} (\bibinfo {year} {2008})}\BibitemShut {NoStop}%
\bibitem [{\citenamefont {Krishnan}\ \emph {et~al.}(1998)\citenamefont
  {Krishnan} \emph {et~al.}}]{krishnan98}%
  \BibitemOpen
  \bibfield  {author} {\bibinfo {author} {\bibfnamefont {A.}~\bibnamefont
  {Krishnan}} \emph {et~al.},\ }\href@noop {} {\bibfield  {journal} {\bibinfo
  {journal} {Phys. Rev. B}\ }\textbf {\bibinfo {volume} {58}},\ \bibinfo
  {pages} {14013} (\bibinfo {year} {1998})}\BibitemShut {NoStop}%
\bibitem [{\citenamefont {Overney}\ \emph {et~al.}(1993)\citenamefont
  {Overney}, \citenamefont {Zhong},\ and\ \citenamefont {Tomanek}}]{overney}%
  \BibitemOpen
  \bibfield  {author} {\bibinfo {author} {\bibfnamefont {G.}~\bibnamefont
  {Overney}}, \bibinfo {author} {\bibfnamefont {W.}~\bibnamefont {Zhong}}, \
  and\ \bibinfo {author} {\bibfnamefont {D.}~\bibnamefont {Tomanek}},\
  }\href@noop {} {\bibfield  {journal} {\bibinfo  {journal} {Z. Phys. D}\
  }\textbf {\bibinfo {volume} {27}},\ \bibinfo {pages} {93} (\bibinfo {year}
  {1993})}\BibitemShut {NoStop}%
\bibitem [{\citenamefont {Huang}\ \emph {et~al.}(2009)\citenamefont {Huang}
  \emph {et~al.}}]{huanga09}%
  \BibitemOpen
  \bibfield  {author} {\bibinfo {author} {\bibfnamefont {M.}~\bibnamefont
  {Huang}} \emph {et~al.},\ }\href@noop {} {\bibfield  {journal} {\bibinfo
  {journal} {Proc. Natl. Acad. Sci. U.S.A.}\ }\textbf {\bibinfo {volume}
  {106}},\ \bibinfo {pages} {7304} (\bibinfo {year} {2009})}\BibitemShut
  {NoStop}%
\bibitem [{\citenamefont {Gui}\ \emph {et~al.}(2008)\citenamefont {Gui},
  \citenamefont {Li},\ and\ \citenamefont {Zhong}}]{gui08}%
  \BibitemOpen
  \bibfield  {author} {\bibinfo {author} {\bibfnamefont {G.}~\bibnamefont
  {Gui}}, \bibinfo {author} {\bibfnamefont {J.}~\bibnamefont {Li}}, \ and\
  \bibinfo {author} {\bibfnamefont {J.}~\bibnamefont {Zhong}},\ }\href@noop {}
  {\bibfield  {journal} {\bibinfo  {journal} {Phys. Rev. B}\ }\textbf {\bibinfo
  {volume} {78}},\ \bibinfo {pages} {075435} (\bibinfo {year}
  {2008})}\BibitemShut {NoStop}%
\bibitem [{\citenamefont {Pereira}\ \emph {et~al.}(2009)\citenamefont
  {Pereira}, \citenamefont {Neto},\ and\ \citenamefont {Peres}}]{pereira09}%
  \BibitemOpen
  \bibfield  {author} {\bibinfo {author} {\bibfnamefont {V.~M.}\ \bibnamefont
  {Pereira}}, \bibinfo {author} {\bibfnamefont {A.~H.~C.}\ \bibnamefont
  {Neto}}, \ and\ \bibinfo {author} {\bibfnamefont {N.~M.~R.}\ \bibnamefont
  {Peres}},\ }\href@noop {} {\bibfield  {journal} {\bibinfo  {journal} {Phys
  Rev B}\ }\textbf {\bibinfo {volume} {80}},\ \bibinfo {pages} {045401}
  (\bibinfo {year} {2009})}\BibitemShut {NoStop}%
\bibitem [{\citenamefont {Cocco}\ \emph {et~al.}(2010)\citenamefont {Cocco},
  \citenamefont {Cadelano},\ and\ \citenamefont {Colombo}}]{colombo}%
  \BibitemOpen
  \bibfield  {author} {\bibinfo {author} {\bibfnamefont {G.}~\bibnamefont
  {Cocco}}, \bibinfo {author} {\bibfnamefont {E.}~\bibnamefont {Cadelano}}, \
  and\ \bibinfo {author} {\bibfnamefont {L.}~\bibnamefont {Colombo}},\
  }\href@noop {} {\bibfield  {journal} {\bibinfo  {journal} {Phys. Rev. B}\
  }\textbf {\bibinfo {volume} {81}},\ \bibinfo {pages} {241412R} (\bibinfo
  {year} {2010})}\BibitemShut {NoStop}%
\bibitem [{\citenamefont {Fasolino}\ \emph {et~al.}(2007)\citenamefont
  {Fasolino}, \citenamefont {Los},\ and\ \citenamefont
  {Katsnelson}}]{fasolino07}%
  \BibitemOpen
  \bibfield  {author} {\bibinfo {author} {\bibfnamefont {A.}~\bibnamefont
  {Fasolino}}, \bibinfo {author} {\bibfnamefont {J.}~\bibnamefont {Los}}, \
  and\ \bibinfo {author} {\bibfnamefont {M.~I.}\ \bibnamefont {Katsnelson}},\
  }\href@noop {} {\bibfield  {journal} {\bibinfo  {journal} {Nat. Mater.}\
  }\textbf {\bibinfo {volume} {6}},\ \bibinfo {pages} {858.} (\bibinfo {year}
  {2007})}\BibitemShut {NoStop}%
\bibitem [{\citenamefont {Boukhalov}\ and\ \citenamefont
  {Katsnelson}(2008)}]{meyer07}%
  \BibitemOpen
  \bibfield  {author} {\bibinfo {author} {\bibfnamefont {D.}~\bibnamefont
  {Boukhalov}}\ and\ \bibinfo {author} {\bibfnamefont {M.~I.}\ \bibnamefont
  {Katsnelson}},\ }\href@noop {} {\bibfield  {journal} {\bibinfo  {journal} {J.
  Am. Soc.}\ }\textbf {\bibinfo {volume} {130}},\ \bibinfo {pages} {10697}
  (\bibinfo {year} {2008})}\BibitemShut {NoStop}%
\bibitem [{\citenamefont {Stolyarova}\ \emph {et~al.}(2007)\citenamefont
  {Stolyarova} \emph {et~al.}}]{stolyarova07}%
  \BibitemOpen
  \bibfield  {author} {\bibinfo {author} {\bibfnamefont {E.}~\bibnamefont
  {Stolyarova}} \emph {et~al.},\ }\href@noop {} {\bibfield  {journal} {\bibinfo
   {journal} {Proc. Natl. Acad. Sci. U.S.A.}\ }\textbf {\bibinfo {volume}
  {104}},\ \bibinfo {pages} {9209} (\bibinfo {year} {2007})}\BibitemShut
  {NoStop}%
\bibitem [{\citenamefont {Ishigami}\ \emph {et~al.}(2007)\citenamefont
  {Ishigami} \emph {et~al.}}]{ishigami07}%
  \BibitemOpen
  \bibfield  {author} {\bibinfo {author} {\bibfnamefont {M.}~\bibnamefont
  {Ishigami}} \emph {et~al.},\ }\href@noop {} {\bibfield  {journal} {\bibinfo
  {journal} {Nano Lett.}\ }\textbf {\bibinfo {volume} {7}},\ \bibinfo {pages}
  {1643.} (\bibinfo {year} {2007})}\BibitemShut {NoStop}%
\bibitem [{\citenamefont {Guinea}\ \emph {et~al.}(2008)\citenamefont {Guinea},
  \citenamefont {Katsnelson},\ and\ \citenamefont {Vozmediano}}]{guinea08}%
  \BibitemOpen
  \bibfield  {author} {\bibinfo {author} {\bibfnamefont {F.}~\bibnamefont
  {Guinea}}, \bibinfo {author} {\bibfnamefont {M.~I.}\ \bibnamefont
  {Katsnelson}}, \ and\ \bibinfo {author} {\bibfnamefont {M.~A.~H.}\
  \bibnamefont {Vozmediano}},\ }\href@noop {} {\bibfield  {journal} {\bibinfo
  {journal} {Phys. Rev. B}\ }\textbf {\bibinfo {volume} {77}},\ \bibinfo
  {pages} {075422} (\bibinfo {year} {2008})}\BibitemShut {NoStop}%
\bibitem [{\citenamefont {Wehling}\ \emph {et~al.}(2008)\citenamefont {Wehling}
  \emph {et~al.}}]{wehling08}%
  \BibitemOpen
  \bibfield  {author} {\bibinfo {author} {\bibfnamefont {T.~O.}\ \bibnamefont
  {Wehling}} \emph {et~al.},\ }\href@noop {} {\bibfield  {journal} {\bibinfo
  {journal} {Eur. Phys. Lett.}\ }\textbf {\bibinfo {volume} {84}},\ \bibinfo
  {pages} {17003} (\bibinfo {year} {2008})}\BibitemShut {NoStop}%
\bibitem [{\citenamefont {Schniepp}\ \emph {et~al.}(2008)\citenamefont
  {Schniepp} \emph {et~al.}}]{Car08}%
  \BibitemOpen
  \bibfield  {author} {\bibinfo {author} {\bibfnamefont {H.~C.}\ \bibnamefont
  {Schniepp}} \emph {et~al.},\ }\href@noop {} {\bibfield  {journal} {\bibinfo
  {journal} {ACS Nano}\ }\textbf {\bibinfo {volume} {2}},\ \bibinfo {pages}
  {2577} (\bibinfo {year} {2008})}\BibitemShut {NoStop}%
\bibitem [{\citenamefont {Bangert}\ \emph {et~al.}(2009)\citenamefont {Bangert}
  \emph {et~al.}}]{Bangert09}%
  \BibitemOpen
  \bibfield  {author} {\bibinfo {author} {\bibfnamefont {U.}~\bibnamefont
  {Bangert}} \emph {et~al.},\ }\href@noop {} {\bibfield  {journal} {\bibinfo
  {journal} {Phys. Stat. Sol. A}\ }\textbf {\bibinfo {volume} {206}},\ \bibinfo
  {pages} {1117} (\bibinfo {year} {2009})}\BibitemShut {NoStop}%
\bibitem [{\citenamefont {Thompson-Flagg}\ \emph {et~al.}(2009)\citenamefont
  {Thompson-Flagg}, \citenamefont {Moura},\ and\ \citenamefont
  {Marder}}]{thompson09}%
  \BibitemOpen
  \bibfield  {author} {\bibinfo {author} {\bibfnamefont {R.~C.}\ \bibnamefont
  {Thompson-Flagg}}, \bibinfo {author} {\bibfnamefont {M.~J.~B.}\ \bibnamefont
  {Moura}}, \ and\ \bibinfo {author} {\bibfnamefont {M.}~\bibnamefont
  {Marder}},\ }\href@noop {} {\bibfield  {journal} {\bibinfo  {journal} {Eur.
  Phys. Lett.}\ }\textbf {\bibinfo {volume} {85}},\ \bibinfo {pages} {46002.}
  (\bibinfo {year} {2009})}\BibitemShut {NoStop}%
\bibitem [{\citenamefont {G\'omez-Navarro}\ \emph {et~al.}(2010)\citenamefont
  {G\'omez-Navarro} \emph {et~al.}}]{kern}%
  \BibitemOpen
  \bibfield  {author} {\bibinfo {author} {\bibfnamefont {C.}~\bibnamefont
  {G\'omez-Navarro}} \emph {et~al.},\ }\href@noop {} {\bibfield  {journal}
  {\bibinfo  {journal} {Nano Lett.}\ }\textbf {\bibinfo {volume} {10}},\
  \bibinfo {pages} {1144} (\bibinfo {year} {2010})}\BibitemShut {NoStop}%
\bibitem [{\citenamefont {Santos}\ \emph {et~al.}(2008)\citenamefont {Santos}
  \emph {et~al.}}]{santos08}%
  \BibitemOpen
  \bibfield  {author} {\bibinfo {author} {\bibfnamefont {E.~J.~G.}\
  \bibnamefont {Santos}} \emph {et~al.},\ }\href@noop {} {\bibfield  {journal}
  {\bibinfo  {journal} {Phys. Rev. B}\ }\textbf {\bibinfo {volume} {78}},\
  \bibinfo {pages} {195420} (\bibinfo {year} {2008})}\BibitemShut {NoStop}%
\bibitem [{\citenamefont {Santos}\ \emph
  {et~al.}(2010{\natexlab{a}})\citenamefont {Santos}, \citenamefont
  {S\'anchez-Portal},\ and\ \citenamefont {Ayuela}}]{santos10a}%
  \BibitemOpen
  \bibfield  {author} {\bibinfo {author} {\bibfnamefont {E.~J.~G.}\
  \bibnamefont {Santos}}, \bibinfo {author} {\bibfnamefont {D.}~\bibnamefont
  {S\'anchez-Portal}}, \ and\ \bibinfo {author} {\bibfnamefont
  {A.}~\bibnamefont {Ayuela}},\ }\href@noop {} {\bibfield  {journal} {\bibinfo
  {journal} {Phys. Rev. B}\ }\textbf {\bibinfo {volume} {81}},\ \bibinfo
  {pages} {125433} (\bibinfo {year} {2010}{\natexlab{a}})}\BibitemShut
  {NoStop}%
\bibitem [{\citenamefont {Santos}\ \emph
  {et~al.}(2010{\natexlab{b}})\citenamefont {Santos}, \citenamefont {Ayuela},\
  and\ \citenamefont {S\'anchez-Portal}}]{santos10b}%
  \BibitemOpen
  \bibfield  {author} {\bibinfo {author} {\bibfnamefont {E.~J.~G.}\
  \bibnamefont {Santos}}, \bibinfo {author} {\bibfnamefont {A.}~\bibnamefont
  {Ayuela}}, \ and\ \bibinfo {author} {\bibfnamefont {D.}~\bibnamefont
  {S\'anchez-Portal}},\ }\href@noop {} {\bibfield  {journal} {\bibinfo
  {journal} {New J. Phys.}\ }\textbf {\bibinfo {volume} {12}},\ \bibinfo
  {pages} {053012} (\bibinfo {year} {2010}{\natexlab{b}})}\BibitemShut
  {NoStop}%
\bibitem [{\citenamefont {Lee}\ \emph {et~al.}(1997)\citenamefont {Lee},
  \citenamefont {Kim},\ and\ \citenamefont {Tom\'anek}}]{lee97}%
  \BibitemOpen
  \bibfield  {author} {\bibinfo {author} {\bibfnamefont {Y.}~\bibnamefont
  {Lee}}, \bibinfo {author} {\bibfnamefont {S.}~\bibnamefont {Kim}}, \ and\
  \bibinfo {author} {\bibfnamefont {D.}~\bibnamefont {Tom\'anek}},\ }\href@noop
  {} {\bibfield  {journal} {\bibinfo  {journal} {Phys. Rev. Lett.}\ }\textbf
  {\bibinfo {volume} {78}},\ \bibinfo {pages} {2393} (\bibinfo {year}
  {1997})}\BibitemShut {NoStop}%
\bibitem [{\citenamefont {Ma}\ \emph {et~al.}(2004)\citenamefont {Ma} \emph
  {et~al.}}]{lehtinen}%
  \BibitemOpen
  \bibfield  {author} {\bibinfo {author} {\bibfnamefont {Y.}~\bibnamefont {Ma}}
  \emph {et~al.},\ }\href@noop {} {\bibfield  {journal} {\bibinfo  {journal}
  {New J. Phys.}\ }\textbf {\bibinfo {volume} {6}},\ \bibinfo {pages} {68}
  (\bibinfo {year} {2004})}\BibitemShut {NoStop}%
\bibitem [{\citenamefont {Lehtinen}\ \emph {et~al.}(2003)\citenamefont
  {Lehtinen} \emph {et~al.}}]{lehtinen03}%
  \BibitemOpen
  \bibfield  {author} {\bibinfo {author} {\bibfnamefont {P.}~\bibnamefont
  {Lehtinen}} \emph {et~al.},\ }\href@noop {} {\bibfield  {journal} {\bibinfo
  {journal} {Phys. Rev. Lett.}\ }\textbf {\bibinfo {volume} {91}},\ \bibinfo
  {pages} {017202} (\bibinfo {year} {2003})}\BibitemShut {NoStop}%
\bibitem [{\citenamefont {Esquinazi}\ \emph {et~al.}(2003)\citenamefont
  {Esquinazi} \emph {et~al.}}]{esquinazi03}%
  \BibitemOpen
  \bibfield  {author} {\bibinfo {author} {\bibfnamefont {P.}~\bibnamefont
  {Esquinazi}} \emph {et~al.},\ }\href@noop {} {\bibfield  {journal} {\bibinfo
  {journal} {Phys. Rev. Lett.}\ }\textbf {\bibinfo {volume} {91}},\ \bibinfo
  {pages} {227201} (\bibinfo {year} {2003})}\BibitemShut {NoStop}%
\bibitem [{\citenamefont {Ohldag}\ \emph {et~al.}(2007)\citenamefont {Ohldag}
  \emph {et~al.}}]{ohldag07}%
  \BibitemOpen
  \bibfield  {author} {\bibinfo {author} {\bibfnamefont {H.}~\bibnamefont
  {Ohldag}} \emph {et~al.},\ }\href@noop {} {\ \textbf {\bibinfo {volume}
  {98}},\ \bibinfo {pages} {187204} (\bibinfo {year} {2007})}\BibitemShut
  {NoStop}%
\bibitem [{\citenamefont {Krasheninnikov}\ and\ \citenamefont
  {Banhart}(2007)}]{krasheninnikov07}%
  \BibitemOpen
  \bibfield  {author} {\bibinfo {author} {\bibfnamefont {A.~V.}\ \bibnamefont
  {Krasheninnikov}}\ and\ \bibinfo {author} {\bibfnamefont {F.}~\bibnamefont
  {Banhart}},\ }\href@noop {} {\bibfield  {journal} {\bibinfo  {journal} {Nat.
  Mater.}\ }\textbf {\bibinfo {volume} {6}},\ \bibinfo {pages} {723} (\bibinfo
  {year} {2007})}\BibitemShut {NoStop}%
\bibitem [{\citenamefont {Gomez-Navarro}\ \emph {et~al.}(2005)\citenamefont
  {Gomez-Navarro} \emph {et~al.}}]{gomez05}%
  \BibitemOpen
  \bibfield  {author} {\bibinfo {author} {\bibfnamefont {C.}~\bibnamefont
  {Gomez-Navarro}} \emph {et~al.},\ }\href@noop {} {\bibfield  {journal}
  {\bibinfo  {journal} {Nature Materials}\ }\textbf {\bibinfo {volume} {4}},\
  \bibinfo {pages} {534} (\bibinfo {year} {2005})}\BibitemShut {NoStop}%
\bibitem [{\citenamefont {Ugeda}\ \emph {et~al.}(2010)\citenamefont {Ugeda}
  \emph {et~al.}}]{ugeda}%
  \BibitemOpen
  \bibfield  {author} {\bibinfo {author} {\bibfnamefont {M.~M.}\ \bibnamefont
  {Ugeda}} \emph {et~al.},\ }\href@noop {} {\bibfield  {journal} {\bibinfo
  {journal} {Phys. Rev. Lett.}\ }\textbf {\bibinfo {volume} {104}},\ \bibinfo
  {pages} {096804} (\bibinfo {year} {2010})}\BibitemShut {NoStop}%
\bibitem [{\citenamefont {Pereira}\ \emph {et~al.}(2006)\citenamefont {Pereira}
  \emph {et~al.}}]{pereira06}%
  \BibitemOpen
  \bibfield  {author} {\bibinfo {author} {\bibfnamefont {V.~M.}\ \bibnamefont
  {Pereira}} \emph {et~al.},\ }\href@noop {} {\bibfield  {journal} {\bibinfo
  {journal} {Phys. Rev. Lett.}\ }\textbf {\bibinfo {volume} {96}},\ \bibinfo
  {pages} {036801} (\bibinfo {year} {2006})}\BibitemShut {NoStop}%
\bibitem [{\citenamefont {Yazyev}(1998)}]{yazyev98}%
  \BibitemOpen
  \bibfield  {author} {\bibinfo {author} {\bibfnamefont {O.~V.}\ \bibnamefont
  {Yazyev}},\ }\href@noop {} {\bibfield  {journal} {\bibinfo  {journal} {Phys.
  Rev. Lett.}\ }\textbf {\bibinfo {volume} {101}},\ \bibinfo {pages} {037203}
  (\bibinfo {year} {1998})}\BibitemShut {NoStop}%
\bibitem [{\citenamefont {Fujita}\ \emph {et~al.}(1996)\citenamefont {Fujita}
  \emph {et~al.}}]{fujita96}%
  \BibitemOpen
  \bibfield  {author} {\bibinfo {author} {\bibfnamefont {M.}~\bibnamefont
  {Fujita}} \emph {et~al.},\ }\href@noop {} {\bibfield  {journal} {\bibinfo
  {journal} {Phys. Soc. Jap.}\ }\textbf {\bibinfo {volume} {65}},\ \bibinfo
  {pages} {1920.} (\bibinfo {year} {1996})}\BibitemShut {NoStop}%
\bibitem [{\citenamefont {Enoki}\ \emph {et~al.}(2007)\citenamefont {Enoki},
  \citenamefont {Kobayashi},\ and\ \citenamefont {Fukui}}]{enoki07}%
  \BibitemOpen
  \bibfield  {author} {\bibinfo {author} {\bibfnamefont {T.}~\bibnamefont
  {Enoki}}, \bibinfo {author} {\bibfnamefont {Y.}~\bibnamefont {Kobayashi}}, \
  and\ \bibinfo {author} {\bibfnamefont {K.~I.}\ \bibnamefont {Fukui}},\
  }\href@noop {} {\bibfield  {journal} {\bibinfo  {journal} {Int. Rev. Phys.
  Chem.}\ }\textbf {\bibinfo {volume} {26}},\ \bibinfo {pages} {609} (\bibinfo
  {year} {2007})}\BibitemShut {NoStop}%
\bibitem [{\citenamefont {Meyer}\ \emph {et~al.}(2007)\citenamefont {Meyer}
  \emph {et~al.}}]{ripples07}%
  \BibitemOpen
  \bibfield  {author} {\bibinfo {author} {\bibfnamefont {J.~C.}\ \bibnamefont
  {Meyer}} \emph {et~al.},\ }\href@noop {} {\bibfield  {journal} {\bibinfo
  {journal} {Nature}\ }\textbf {\bibinfo {volume} {446}},\ \bibinfo {pages}
  {60.} (\bibinfo {year} {2007})}\BibitemShut {NoStop}%
\bibitem [{\citenamefont {de~Parga}\ \emph {et~al.}(2008)\citenamefont
  {de~Parga} \emph {et~al.}}]{Vazquez08}%
  \BibitemOpen
  \bibfield  {author} {\bibinfo {author} {\bibfnamefont {A.~L.~V.}\
  \bibnamefont {de~Parga}} \emph {et~al.},\ }\href@noop {} {\bibfield
  {journal} {\bibinfo  {journal} {Phys. Rev. Lett.}\ }\textbf {\bibinfo
  {volume} {100}},\ \bibinfo {pages} {056807} (\bibinfo {year}
  {2008})}\BibitemShut {NoStop}%
\bibitem [{\citenamefont {Geringer}\ \emph {et~al.}(2009)\citenamefont
  {Geringer} \emph {et~al.}}]{Morgenstern09}%
  \BibitemOpen
  \bibfield  {author} {\bibinfo {author} {\bibfnamefont {V.}~\bibnamefont
  {Geringer}} \emph {et~al.},\ }\href@noop {} {\bibfield  {journal} {\bibinfo
  {journal} {Phys. Rev. Lett.}\ }\textbf {\bibinfo {volume} {102}},\ \bibinfo
  {pages} {076102} (\bibinfo {year} {2009})}\BibitemShut {NoStop}%
\bibitem [{\citenamefont {Amara1}\ \emph {et~al.}()\citenamefont {Amara1} \emph
  {et~al.}}]{amara07}%
  \BibitemOpen
  \bibfield  {author} {\bibinfo {author} {\bibfnamefont {H.}~\bibnamefont
  {Amara1}} \emph {et~al.},\ }\href@noop {} {\ }\BibitemShut {NoStop}%
\bibitem [{\citenamefont {El-Barbary}\ \emph {et~al.}(2003)\citenamefont
  {El-Barbary} \emph {et~al.}}]{ewels03}%
  \BibitemOpen
  \bibfield  {author} {\bibinfo {author} {\bibfnamefont {A.~A.}\ \bibnamefont
  {El-Barbary}} \emph {et~al.},\ }\href@noop {} {\bibfield  {journal} {\bibinfo
   {journal} {Phys. Rev. B}\ }\textbf {\bibinfo {volume} {68}},\ \bibinfo
  {pages} {144107} (\bibinfo {year} {2003})}\BibitemShut {NoStop}%
\bibitem [{\citenamefont {R\"{o}hrl}\ \emph {et~al.}(2008)\citenamefont
  {R\"{o}hrl} \emph {et~al.}}]{strainAPL}%
  \BibitemOpen
  \bibfield  {author} {\bibinfo {author} {\bibfnamefont {J.}~\bibnamefont
  {R\"{o}hrl}} \emph {et~al.},\ }\href@noop {} {\bibfield  {journal} {\bibinfo
  {journal} {Appl. Phys. Lett.}\ }\textbf {\bibinfo {volume} {92}},\ \bibinfo
  {pages} {201918} (\bibinfo {year} {2008})}\BibitemShut {NoStop}%
\bibitem [{\citenamefont {Rutter}\ \emph {et~al.}(2007)\citenamefont {Rutter}
  \emph {et~al.}}]{Stroscio07}%
  \BibitemOpen
  \bibfield  {author} {\bibinfo {author} {\bibfnamefont {G.~M.}\ \bibnamefont
  {Rutter}} \emph {et~al.},\ }\href@noop {} {\bibfield  {journal} {\bibinfo
  {journal} {Phys. Rev. B}\ }\textbf {\bibinfo {volume} {76}},\ \bibinfo
  {pages} {235416} (\bibinfo {year} {2007})}\BibitemShut {NoStop}%
\bibitem [{\citenamefont {Mohiuddin}\ \emph {et~al.}(2009)\citenamefont
  {Mohiuddin} \emph {et~al.}}]{Mohiuddin09}%
  \BibitemOpen
  \bibfield  {author} {\bibinfo {author} {\bibfnamefont {T.~M.~G.}\
  \bibnamefont {Mohiuddin}} \emph {et~al.},\ }\href@noop {} {\bibfield
  {journal} {\bibinfo  {journal} {Phys. Rev. B}\ }\textbf {\bibinfo {volume}
  {79}},\ \bibinfo {pages} {205433} (\bibinfo {year} {2009})}\BibitemShut
  {NoStop}%
\bibitem [{\citenamefont {Kim}\ \emph {et~al.}(2009)\citenamefont {Kim} \emph
  {et~al.}}]{Kim09}%
  \BibitemOpen
  \bibfield  {author} {\bibinfo {author} {\bibfnamefont {K.~S.}\ \bibnamefont
  {Kim}} \emph {et~al.},\ }\href@noop {} {\bibfield  {journal} {\bibinfo
  {journal} {Nature}\ }\textbf {\bibinfo {volume} {457}},\ \bibinfo {pages}
  {706} (\bibinfo {year} {2009})}\BibitemShut {NoStop}%
\bibitem [{\citenamefont {Rodriguez-Manzo}\ and\ \citenamefont
  {Banhart}(2009)}]{Banhart09}%
  \BibitemOpen
  \bibfield  {author} {\bibinfo {author} {\bibfnamefont {J.~A.}\ \bibnamefont
  {Rodriguez-Manzo}}\ and\ \bibinfo {author} {\bibfnamefont {F.}~\bibnamefont
  {Banhart}},\ }\href@noop {} {\bibfield  {journal} {\bibinfo  {journal} {Nano
  Lett.}\ }\textbf {\bibinfo {volume} {9}},\ \bibinfo {pages} {2285} (\bibinfo
  {year} {2009})}\BibitemShut {NoStop}%
\bibitem [{\citenamefont {Wiesendanger}(2009)}]{Wiesendanger09}%
  \BibitemOpen
  \bibfield  {author} {\bibinfo {author} {\bibfnamefont {R.}~\bibnamefont
  {Wiesendanger}},\ }\href@noop {} {\bibfield  {journal} {\bibinfo  {journal}
  {Rev. Mod. Phys.}\ }\textbf {\bibinfo {volume} {81}},\ \bibinfo {pages}
  {1495} (\bibinfo {year} {2009})}\BibitemShut {NoStop}%
\bibitem [{\citenamefont {Kohn}\ and\ \citenamefont {Sham}(1965)}]{kohn1965}%
  \BibitemOpen
  \bibfield  {author} {\bibinfo {author} {\bibfnamefont {W.}~\bibnamefont
  {Kohn}}\ and\ \bibinfo {author} {\bibfnamefont {L.~J.}\ \bibnamefont
  {Sham}},\ }\href@noop {} {\bibfield  {journal} {\bibinfo  {journal} {Phys.
  Rev.}\ }\textbf {\bibinfo {volume} {140}},\ \bibinfo {pages} {A1133}
  (\bibinfo {year} {1965})}\BibitemShut {NoStop}%
\bibitem [{\citenamefont {Soler}\ \emph {et~al.}(2002)\citenamefont {Soler}
  \emph {et~al.}}]{soler02}%
  \BibitemOpen
  \bibfield  {author} {\bibinfo {author} {\bibfnamefont {J.~M.}\ \bibnamefont
  {Soler}} \emph {et~al.},\ }\href@noop {} {\bibfield  {journal} {\bibinfo
  {journal} {J. Phys.: Conden. Matter}\ }\textbf {\bibinfo {volume} {14}},\
  \bibinfo {pages} {2745} (\bibinfo {year} {2002})}\BibitemShut {NoStop}%
\bibitem [{\citenamefont {Perdew}\ \emph {et~al.}(1996)\citenamefont {Perdew},
  \citenamefont {Burke},\ and\ \citenamefont {Ernzerhof}}]{gga}%
  \BibitemOpen
  \bibfield  {author} {\bibinfo {author} {\bibfnamefont {J.~P.}\ \bibnamefont
  {Perdew}}, \bibinfo {author} {\bibfnamefont {K.}~\bibnamefont {Burke}}, \
  and\ \bibinfo {author} {\bibfnamefont {M.}~\bibnamefont {Ernzerhof}},\
  }\href@noop {} {\bibfield  {journal} {\bibinfo  {journal} {Phys. Rev. Lett.}\
  }\textbf {\bibinfo {volume} {77}},\ \bibinfo {pages} {3865.} (\bibinfo {year}
  {1996})}\BibitemShut {NoStop}%
\bibitem [{\citenamefont {Troullier}\ and\ \citenamefont
  {Martins}(1991)}]{troullier91}%
  \BibitemOpen
  \bibfield  {author} {\bibinfo {author} {\bibfnamefont {N.}~\bibnamefont
  {Troullier}}\ and\ \bibinfo {author} {\bibfnamefont {J.~L.}\ \bibnamefont
  {Martins}},\ }\href@noop {} {\bibfield  {journal} {\bibinfo  {journal} {Phys.
  Rev. B}\ }\textbf {\bibinfo {volume} {43}},\ \bibinfo {pages} {1993}
  (\bibinfo {year} {1991})}\BibitemShut {NoStop}%
\bibitem [{\citenamefont {Monkhosrt}\ and\ \citenamefont
  {Pack}(1976)}]{monkhosrt76}%
  \BibitemOpen
  \bibfield  {author} {\bibinfo {author} {\bibfnamefont {H.~J.}\ \bibnamefont
  {Monkhosrt}}\ and\ \bibinfo {author} {\bibfnamefont {J.~D.}\ \bibnamefont
  {Pack}},\ }\href@noop {} {\bibfield  {journal} {\bibinfo  {journal} {Phys.
  Rev. B}\ }\textbf {\bibinfo {volume} {13}},\ \bibinfo {pages} {5188}
  (\bibinfo {year} {1976})}\BibitemShut {NoStop}%
\end{thebibliography}%
\end{document}